\documentstyle[aps, 12pt]{revtex}
\title{Inelastic Neutron scattering in $\bf CeSi_{2-x}Ga_x$ ferromagnetic
Kondo lattice compounds }
\author{K. R. Priolkar$^1$, Mala N. Rao$^2$, R. B. Prabhu$^1$, P. R. Sarode$^1$,
S. K. Paranjpe$^2$, P. Raj$^3$ and A. Sathyamoorthy$^3$}
\address{$^1$ Department of Physics, Goa University, Taleigao Plateau, 
Goa 403 205, India.}
\address{$^2$ Solid State Physics Division, Bhabha Atomic Research
Centre,Trombay, Bombay, 400 085, India.}
\address{$^3$ Chemistry Division, Bhabha Atomic Research Centre, Trombay, 
Bombay, 400 085, India.}
\date{\today}
\begin{document}
\maketitle
\vspace{1cm}
\begin{abstract}
Inelastic neutron scattering investigation on ferromagnetic Kondo lattice
compounds belonging to $\rm CeSi_{2-x}Ga_x$, x = 0.7, 1.0 and 1.3, system is
reported.The thermal evolution of the quasielastic response shows that the
Kondo interactions dominate over the RKKY interactions with increase in Ga
concentration from 0.7 to 1.3. This is related to the increase in k-f
hybridization with increasing Ga concentration. The high energy response
indicates the ground state to be split by crystal field in all three
compounds. Using the experimental results we have calculated the crystal
field
parameters in all three compounds studied here.
\end{abstract}
\newpage
%
\section{Introduction}
Cerium in its intermetallic compounds and alloys exhibits various types of
anomalous ground states that are closely linked with the hybridization
strength between the conduction electrons and the Ce 4f electrons. The two
processes viz, interatomic RKKY interaction and the single site Kondo
interaction compete with each other in cerium Kondo lattice compounds.
Dominance of one over other leads to either magnetically ordered or a non
magnetic ground state. Many cerium systems \cite{1,2,3,4} have been studied
to
understand the competition and various types of ground states these materials
exhibit. Cerium silicides with either hole doping ($\rm CeSi_x$, \cite{5},
$\rm Ce(SiAl)_2$ \cite{6}, $\rm Ce(SiGa)_2$ \cite{7,8,9}) or isoelectronic
substitution ($\rm Ce(SiGe)_2$ \cite{10} have been of lot of interest in
recent years. All these systems show an evolution from nonmagnetic to a
magnetically ordered ground state. In $\rm CeSi_{2-x}Ga_x$, the system
evolves from nonmagnetic - ferromagnetic - antiferromagnetic ground state
associated with a structural transition from tetragonal - hexagonal structure
near the ferro - antiferro magnetic transition \cite{7,8,9}.

In addition to the competition between the Kondo and RKKY interactions, in
nearly all cerium Kondo systems, crystalline electric field (CEF) plays an
important role in deciding the ground state of the system. It determines the
degeneracy of the f level and the f electron ground state wave function
involved in the hybridization and hence is also required in any rigorous
theoretical model \cite{11}. In addition, Levy and Zhang \cite{12} have also
proposed that the hybridization interaction between the localized f electron
states and the band states plays an important role in the strength of the CF
potential.

The most direct method of determining CEF in a metallic compound is inelastic
neutron scattering for which the scattering cross-section is proportional to
dynamic susceptibility. In $\rm CeSi_2$ \cite{13} the CEF splits the j = 5/2
ground state into three doublets with the excited doublets lying at 297K and
555K from the lowest doublet. Khogi et al \cite{14} have shown that in $\rm
CeSi_x$ system the line widths and the excitation energies of the CEF
doublets scale almost linearly with x and the large linewidths at higher
values of x is due to increasing k-f hybridization with x. In $\rm CeGa_2$
\cite{15} the study of CEF parameters have led to understanding of large easy
plane anisotropy and also the absence of any discontinuity in resistivity
curve at magnetic ordering temperature.

In this paper we report our inelastic neutron scattering studies on three
compounds belonging to $\rm CeSi_{2-x}Ga_x$ with x = 0.7, 1.0 and 1.3. All
these three compositions crystallize in $\alpha \rm-ThSi_2$ type structure
with nearly equal lattice constant values
\cite{7}. Magnetic and transport properties have shown that a strong
competition exists between the intersite RKKY and the single site Kondo
interactions. For x = 0.7 specific heat measurements show a sharp anomaly
indicating stable ferromagnetic order. This peak then broadens out at x = 1.0
and 1.3 which is indicative of the dominance of Kondo interaction over the
RKKY interactions. We have studied the crystal field excitations in these
compounds as well as the thermal evolution of the quasielastic line width in
the temperature range 10 - 100 K.

%
\section{Experiment and Results}
The polycrystalline $\rm CeSi_{2-x}Ga_x$, x = 0.7, 1.0, 1.3 and LaSiGa have
been prepared by arc melting the pure elements in argon atmosphere using the
same procedure as in Ref.7. Neutron diffraction patterns are in good in
agreement with the tetragonal $\alpha \rm-ThSi_2$-type structure and lattice
constant values agreed with those reported in the literature.

Inelastic neutron scattering experiments were performed at the DHRUVA reactor
on the triple axis spectrometer (TAS) installed on a tangential thermal
neutron beam hole T 1007 at Trombay. TAS is a medium resolution spectrometer
which employs Cu (111) plane as monochromator and Si (111) plane as analyzer.
The collimations used are open, 60', 60', and open between reactor and
monochromator, monochromator and sample, sample and analyzer and analyzer and
detector respectively. The spectrometer was operated at fixed final energy,
$E_f$ = 25 meV with incident energy varying from 65 meV to 20 meV at constant
scattering angle, $\phi$. The spectra of each sample were recorded at two
different scattering angles, $\phi$ = 20$^\circ$ and 95$^\circ$ (Q = 1
\AA$^{-1}$ and 5 \AA$^{-1}$) and at different temperatures from 10 K to 100 K
using a closed cycle refrigerator.

The phonon contribution for all the three samples was estimated from LaSiGa
data using the scaling method proposed by A. P. Murani \cite{16}. The
magnetic response then obtained may be related to the dynamic susceptibility
$\chi^{''}(Q, \omega)$.
\begin{equation}
S(Q,\omega) = A \left[ \frac{1}{1-exp(-\hbar\omega/k_BT)} \right] f^2(Q)
\times \chi^{''}(Q, \omega)
\end{equation}
where $A = 1/(2\pi)(\gamma r_e/\mu_B)^2$ which describes the coupling between
the neutron and electron spin. The Kramers-Kronig relation provides a
relationship between $\chi^{''}(Q, \omega)$ and static susceptibility which
can be written as
\begin{equation}
\chi^{''}(Q,\omega) = \pi\hbar\omega\chi(Q)P(Q,\omega)
\end{equation}
The static susceptibility $\chi(Q)$ is related to the bulk susceptibility
$\chi_{bulk}$ via a magnetic form factor $f(Q)$, $\chi(Q) =
f(Q)^2\chi_{bulk}$. $P(Q, \omega)$ is a spectral function which fulfills the
relation $\int_{-\infty}^{\infty} P(Q, \omega) d\omega$ = 1. A lorentzian
form
is usually assumed to describe the relaxation processes. For a pure
quasielastic response the Lorentzians centered at $\hbar\omega = 0$ and in
presence of crystal field splittings $P(Q,\omega)$ is described by a series
of lorentzians centered at $\hbar\omega = 0$ (quasielastic) and
$\pm\hbar\omega_i$ (crystal field excitations) as
\begin{equation}
\chi(Q)P(Q,\omega)~ = ~\frac{A_0(T)\Gamma_0(T)}{\Gamma_0^2(T) + \omega^2} ~+~
\sum_1^n \frac{A_i(T)\Gamma_i(T)}{\Gamma_i^2(T) + (\omega \pm \omega_i)^2}
\end{equation}
where $A_0$, $A_i$ are the amplitudes and $\Gamma_0$, $\Gamma_i$ the half
widths of the quasielastic and inelastic structures respectively.

The normalized spectra measured at 12 K on TAS after phonon correction and
correction for empty cell scattering are shown in Fig. 1 for the three $\rm
CeSi_{2-x}Ga_x$ compounds with x = 0.7, 1.0 and 1.3. The quasielastic peak
and 0 meV energy transfer and the inelastic peaks indicates the presence of
magnetic scattering in these samples. The solid line in the figure is least
square fit line to the data using the equation (3). The least square fit
parameters obtained by fitting the spectra at all temperatures for all the
three compounds are given in Table I. From the table it is clear that in all
the three samples there is quasielastic broadening as well as inelastic peaks
due to CEF splitting of the ground state. In the case of $\rm
CeSi_{1.3}Ga_{0.7}$ only one broad inelastic peak can be seen while for the
other two samples the spectra can be fitted to two lorentzians which implies
that the ground state in split in three doublets as expected for tetragonal
point symmetry. Single inelastic peak has been observed previously
\cite{17,18} in case of tetragonal symmetry which has been explained by
doublet quasiquatret splitting of the ground state. The other reason could
also be that the excited doublets are lying very close to each other and due
to our moderate energy resolution it is not possible to separate them.
%
\section{Discussion}
The residual quasielastic line width, in Kondo and heavy fermion systems,
gives the fluctuation rate and it is often taken as the measure of Kondo
temperature or spin fluctuation temperature. The thermal evolution of the
quasielastic linewidth can either be fitted to $T^{1/2}$ law or to a linear T
dependence in accordance with Korringa law \cite{19,20}. In figure 2 we have
plotted the temperature dependence of the quasielastic line width in all the
three compounds. In case of $\rm CeSi_{1.3}Ga_{0.7}$, both linear as well as
$\sqrt T$ lines fit equally well to the data. The data point corresponding to
line width at 12K is excluded as it is just within the resolution. $T_K$ in
all the three compounds is estimated from $\Gamma = \Gamma_0 + k_BT^{1/2}$
with
$\Gamma_0 = k_BT_K$. For x = 0.7 sample, Kondo temperature is found to be
$T_K$ = 10 K which is in quite good agreement with that deduced from other
techniques. For x = 1.0 also both the equations fit the data equally well
while for x = 1.3 the $\sqrt T$ fit is better. This can be related to the
increasing hybridization with increasing Ga concentration. The values of
$T_K$ obtained for x = 1.0 and 1.3 samples are 17 K and 25 K respectively.
These values are not in agreement with the values of $T_K^l$ deduced from
resistivity data by assuming the values of excitation energies as that of
$\rm CeSi_{1.7}$ \cite{8}. It may be noted here that calculating $\rm T_K^l$
using the actual values of excitation energies gives a far better agreement
with the $T_K$ values deduced from quasielastic line widths.

We have also tried to study the crossover phenomena in $\rm CeSi_{2-x}Ga_x$
by taking into account the present experimental results. In a most
divergent approximation of Coqblin-Schrieffer model, $T_K$ for a Ce system in
crystal field is expressed as
\begin{equation}
T_K = D \left( \frac{D}{T_K + \Delta_1} \right) \left( \frac{D}{T_K +
\Delta_2} \right) exp \left[ - \frac{1}{2|\rho J|} \right]
\end{equation}
where $D$ is the half width of the conduction band, $\rho$ is the density of
states of the conduction electrons at $E_F$ and $J$ is the exchange
interaction constant. Using the $T_K$ values deduced from the quasielastic
line widths and assuming $D \sim$ 10000 K we estimate the coupling constant
$|\rho J|$. The temperature associated with RKKY interactions
is expressed as $T_{RKKY} \sim D|\rho J|^2$. The results of this analysis are
summarized in Table II. These results indicate that there exists a cross over
from RKKY dominated region to a region dominated by Kondo interactions at
around x = 0.7. This crossover is caused by increase of k-f coupling constant
(increase in k-f hybridization) with x over its critical value where the
intrasite Kondo screening dominates the ferromagnetic ordering due to
intersite RKKY interactions. This picture is consistent with the one deduced
from the bulk property measurements \cite{8}.

The high energy response in all the three samples can be interpreted in terms
of crystal field excitations broadened by hybridization. The crystal field
Hamiltonian for cerium in tetragonal symmetry can be written as
\begin{equation}
H_{CEF} = B_2^0O_2^0 + B_4^0O_4^0 + B_4^4O_4^4
\end{equation}
where $O_l^m$ are the Stevens operators and $B_l^m$ are the phenomenological
CF parameters. The values of $O_l^m$'s can be obtained from Hutchings
\cite{21}. Diagonalization of the CEF Hamiltonian gives us the eigen values
and eigen functions and with simple algebra $B_l^m$ can be written as
\begin{eqnarray}
B_2^0 & = & \frac{\Delta_1}{14} \left[\eta^2 - \frac{5}{6} \right] - \frac
{\Delta_2}{21}\\
\nonumber \\
B_4^0 & = & \frac{\Delta_1}{210} \left[\eta^2 - \frac{1}{4} \right] + \frac
{\Delta_2}{420}\\
\nonumber \\
B_4^4 & = & \frac{\Delta_1 \eta}{12} \sqrt{\frac{1}{5} (1 - \eta^2)}
\end{eqnarray}
where $\Delta_1$ and $\Delta_2$ are CF excitation energies and $\eta$ is a
coefficient of the doublet wave function
\begin{equation}
|g.s.> = \eta|\pm5/2> + \sqrt{1-\eta^2} |\pm3/2>
\end{equation}
Fixing $\Delta_1$ and
$\Delta_2$ to experimental values, the CF potential then depends upon the
single parameter $\eta$. This can be determined by fitting simultaneously the
single crystal susceptibility data and the neutron scattering data.
Unfortunately single crystal susceptibility data for $\rm CeSi_{2-x}Ga_{x}$
is
not yet reported hence we have calculated the susceptibility $\chi = M/H$ for
each set of $B_2^0$, $B_4^0$ and $B_4^4$ assuming that $H$ is an external
field of 4kG and magnetization $M$ can be averaged according to $M = (M_c +
2M_a)/3$, where $M_c$ and $M_a$ are the magnetizations for fields along $c$-
axis and the $a$- axis respectively. The estimated $\eta$ and molecular field
constant in paramagnetic phase, for x = 0.7, 1.0 and 1.3 were 0.6223 and
$\ldots$, 0.6428 and $\ldots$ and 0.5892 and $\ldots$ respectively. 
The CF parameters which gave best fits are listed in Table III.  It is to be
noted here that the  second order
crystal field parameter $B_2^0$ at 12K is almost constant for all the three
compounds, whereas the fourth order parameters, $B_4^0$ and $B_4^4$, increase
with increasing x. In fact, their behaviour closely resembles the behaviour of
excitation energies if one compares the values of excitation energies and CF
parameters from Table II and Table III respectively. It may also be noted from
Table I that the line width of crystal field excitations (inelastic peak), in
these compounds, is quite large. Such a behaviour has been observed in the
isostructural $\rm CeSi_x$ compounds \cite{14}. Even for the ferromagnetic,
$\rm CeSi_{1.7}$ the broad inelastic peaks correspond with the anomalous
damping of the spin wave excitations in this compound \cite{22}. A similar
behaviour is seen here in case of $\rm CeSi_{1.3}Ga_{0.7}$ compound. However no
single crystal data on this compound is hiterto reported in literature to check
this possibility. The increase of the linewidth of $\rm CeSi_{2-x}Ga_x$ between
x = 1.0 and 1.3 where the ferromagnetism becomes unstable and the Kondo
behaviour develops can be concluded to be due to the increase in the strength
of the 4f electron - conduction electron hybridization.

%
\section{Conclusions}
The temperature dependence of the quasielastic line widths in $\rm
CeSi_{2-x}Ga_x$, (x = 0.7, 1.0 and 1.3) has been studied. In the
concentration range studied here, there exists a crossover from RKKY
dominated region to one dominated by single site Kondo interactions. the
ground state in all these compounds is CF split and we have determined the f
electron ground state wavefunctions and the phenomenological CEF parameters
for
all the three compounds.
\section*{Acknowledgements}
We [K. R. P., R. B. P. and P. R. S.] would like to acknowledge the financial
support from IUC-DAEF, Indore under the research project IUC/BOM/48. It is
also our pleasure to thank Drs. B. A. Dasannacharya, K. R. Rao and P. R.
Vijayaraghavan for their constant encouragement and cooperation during the
course of this work.

\newpage
\begin{table}
\caption{ }
\begin{tabular}{ccccccccc}
T (K) & $A_0$ & $\Gamma_0$ & $A_1$ & $\Gamma_1$ & $\omega_1$ & $A_2$ &
$\Gamma_2$ & $\omega_2$\\
& (Arb. Units) & meV & (Arb. Units) & meV & meV &(Arb. Units) & meV & meV \\
\hline
\multicolumn{7}{c}{x = 0.7}\\
12 & 3.6 & 2.76 & .99 & 9.23 & 13.36  \\
25 & 3.3 & 3.47 & .82 & 10.76 & 13.45 \\
50 & 3.0 & 4.09 & .69 & 12.44 & 13.34\\
100 & 2.7 & 5.28 & .59 & 13.42 & 13.27\\
\multicolumn{7}{c}{x = 1.0}\\
12 & 4.4 & 3.1 & .90 & 4.2 & 15.43 & .25 & 5.3 & 26.46\\
25 & 4.05 & 3.8 & .75 & 5.4 & 15.24& .19 & 6.4 & 26.52\\
50 & 3.6 & 4.5 & .52 & 6.9 &  15.32& .13 & 8.8 & 26.31 \\
100 & 2.6 & 5.3 & .42 & 8.2 & 15.20& .08 & 10.4 & 26.28\\
\multicolumn{7}{c}{x = 1.3}\\
12 & 3.16 & 3.6 & .81 & 5.2 &9.91 & .20 &  6.2 & 20.43\\
25 & 2.8  & 4.4 & .55 & 7.4 &9.86 & .12 &  8.2 & 20.43\\
50 & 2.4  & 5.1 & .42 & 9.05 &9.84 & .08 & 10.8 & 20.32\\
100 & 1.8 & 6.1 & .32 & 10.5 & 9.75& .05 & 12.4 & 20.26\\
\end{tabular}
\end{table}
\newpage
\begin{table}
\caption{Excitation energy $\Delta$, Kondo temperature $\rm T_K$, c-f
coupling constant $\rm |\rho J|$ and $\rm T_{RKKY}$ for $\rm
CeSi_{2-x}Ga_{x}$}
\begin{tabular}{cccccc}
x&$\Delta_1 (K)$&$\Delta_2$&$T_K(K)$&$|\rho J|$&$T_{RKKY}(K)$\\
\hline

0.7&155&155&10&0.0331&10.95\\
\\
1.0&179&307&17&0.0363&13.18\\
\\
1.3&115&237&29&0.0365&13.52\\
\hline
\end{tabular}
\end{table}
\newpage
\begin{table}
\caption{Crystal field parameters of $\rm CeSi_{2-x}Ga_x$}
\begin{tabular}{cccc}
x & $B_2^0$ & $B_4^0$ & $B_4^4$\\
\hline
0.7 & -1.06 & 0.0405 & 0.243\\
1.0 & -1.72 & 0.0750 & 0.283\\
1.3 & -1.32 & 0.0533 & 0.176\\
\end{tabular}
\end{table}
\newpage
\flushleft {\bf \large Captions to Figures}\\

Fig. 1. Magnetic spectral response of $\rm CeSi_{2-x}Ga_x$ (x = 0.7, 1.0 and
1.3) at T = 12 K and $\phi = 20^{\circ}$.\\

Fig. 2. Thermal evolution of the quasielastic linewidth in $\rm
CeSi_{2-x}Ga_x$ (x = 0.7, 1.0 and 1.3)
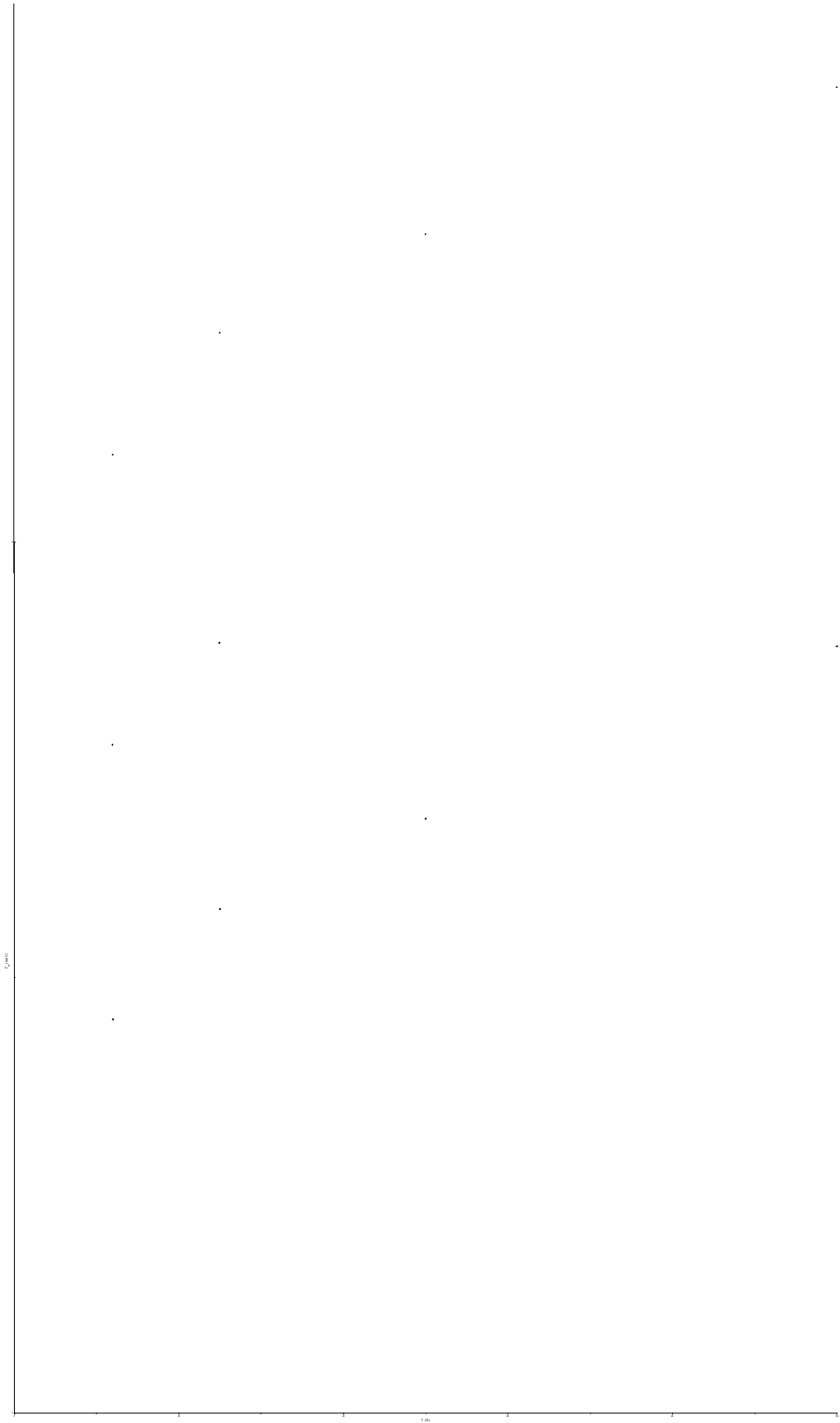
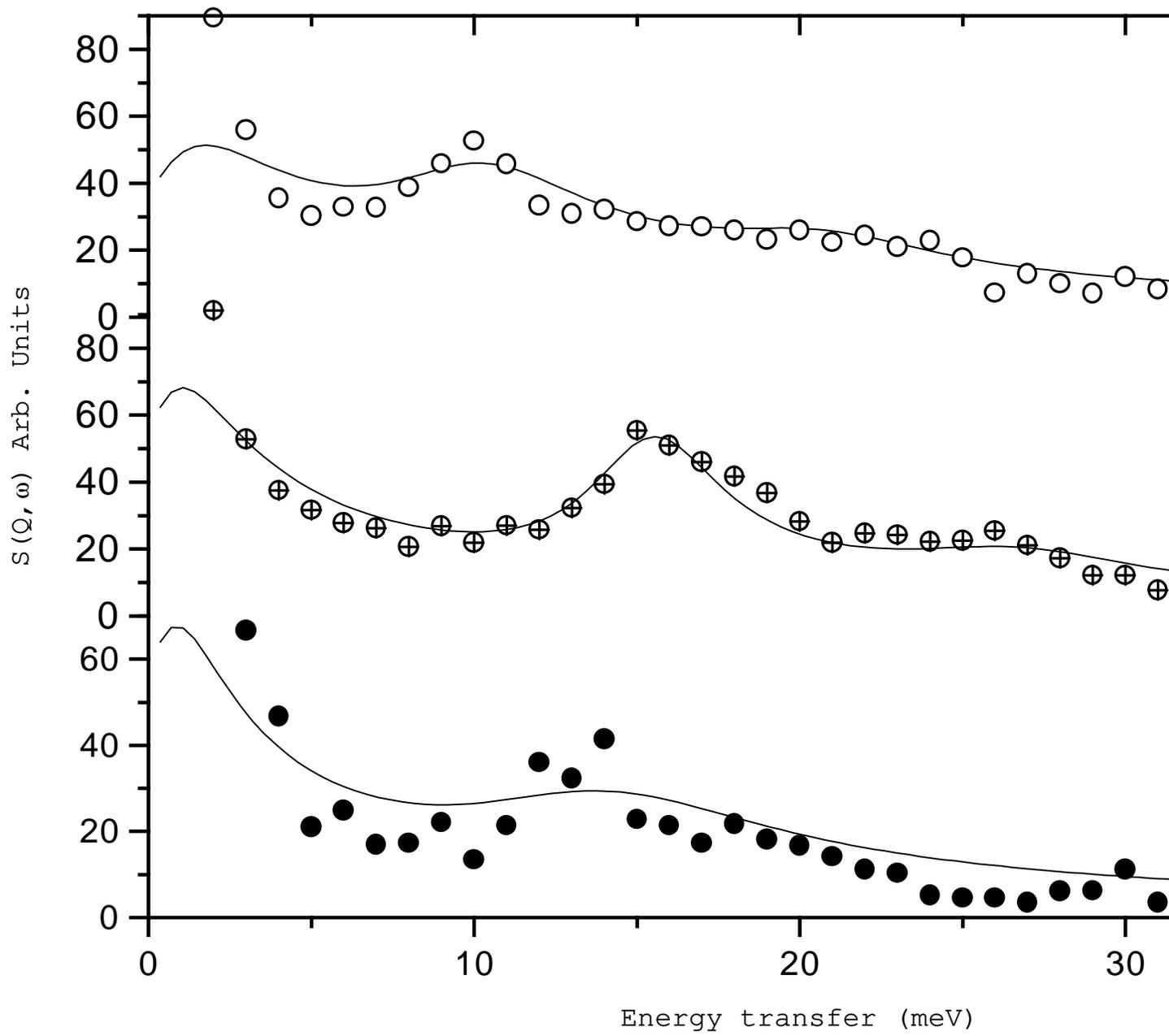
\end{document}